# Gene Algebra from a Genetic Code Algebraic Structure

Robersy Sánchez [1 3*], Eberto Morgado [2] and Ricardo Grau [2 3].

[1] Research Institute of Tropical Roots, Tuber Crops and Banana (INIVIT). Biotechnology group. Santo Domingo. Villa Clara. Cuba.

[2] Faculty of Mathematics Physics and Computation, Central University of Las Villas, Villa Clara, Cuba.

[3] Center of Studies on Informatics, Central University of Las Villas, Villa Clara, Cuba

**Abstract**

The biological distinction between the base positions in the codon, the chemical types of bases (purine and pyrimidine) and their hydrogen bond number have been the most relevant codon properties used in the genetic code analysis. Now, these properties have allowed us to build a Genetic Code ring isomorphic to the ring ($Z_{64}$, +, •) of the integer module 64. On the $Z_{64}$-algebra of the set of $64^N$ codon sequences of length $N$, gene mutations are described by means of endomorphisms $F: (Z_{64})^N \rightarrow (Z_{64})^N$. Endomorphisms and automorphisms helped us describe the gene mutation pathways. For instance, 77.7% mutations in 749 HIV protease gene sequences correspond to unique diagonal endomorphisms of the wild type strain HXB2. In particular, most of the reported mutations that confer drug resistance to the HIV protease gene correspond to diagonal automorphisms of the wild type. What is more, in the human beta-globin gene a similar situation appears where most of the single codon mutations correspond to automorphisms. Hence, in the analyses of molecular evolution process on the DNA sequence set of length $N$, the $Z_{64}$-algebra will help us explain the quantitative relationships between genes.

---

[*] Robersy Sánchez: robersy@uclv.edu.cu

Corresponding address: Apartado postal 697. Santa Clara 1. CP 50100. Villa Clara. Cuba

## 1. Introduction

The genetic code has been represented, usually, in a four-column table where codons are located attending to the second base. Three entries corresponding to the codon bases change their ordering according to different criteria. As a result, hydrophobic and hydrophilic amino acids are found in different columns. Epstein states that amino acids cannot be randomly allocated by just considering the features of the genetic code [10] - fully discussed by Crick [8]- and we believe that the order of codons must reflect their physico-chemical properties. But, could there be an optimal distribution of the table? Gillis et al. have suggested that the genetic code can be optimized to limit translation errors [13]. Thus, there should be an optimal codon order.

We believe that the order of the genetic code must be strongly connected to the physico-chemical properties of the four bases, in particular to the base positions in the codon and to the codon-anticodon interactions [22], where chemical types (purine and pyrimidine) and number of hydrogen bonds have an important role. The importance of the base position is suggested by the error frequency found in the codons. Errors on the third base are more frequent than on the first base, and, in turn, these are more frequent than errors on the second base [35, 11, 27]. These positions, however, are too conservative with respect to changes in polarity of the coded amino acids [1].

Many attempts have been made to introduce a formal characterization of the genetic code [6, 5, 32, 17, 3, 18, 4, 20, 2, 36, 16]. Many of these models highlight a quantitative relationship between codons expressed through the hydrogen bonds and the chemical types of bases. Eventually, these formal descriptions suggest that the hydrogen bond number and the chemical type should be enough to obtain a "natural order" in the codon set. Recently, using these elemental properties of bases Sánchez et al. [30, 31] proposed a Boolean structure of the genetic code in which the partial order of the codon set and the Boolean deductions between codons are connected to the physico-chemical properties of amino acids. Here, nevertheless, we would like to show that by using the same base properties, it is possible to infer a different codon order and a different algebraic structure of the genetic code. This structure will reflect the quantitative relationships between codons in genes.

These relationships are suggested by the codons usage found in genes [24]. In all living organisms, note that some amino acids and some codons are more frequent than others (see http://www.kazusa.or.jp/codon). Each organism has its own "preferred" or more frequently used codons for a given amino acid and their usage is frequent, a tendency called codon bias. For all life forms, codon usage is non-random [12] and associated to various factors such as gene expression level [23], gene length [9] and secondary protein structures [25, 34, 15, 14]. Moreover, most amino acids in all species bear a highly significant association with gene functions, indicating that, in general, codon usage at the level of individual amino acids is closely coordinated with the gene function [12]. This explains why different quantitative codon values may be expressed by codons in genes.

Let's look at a simple, minimal mathematical model to help us better understand the logical framework that underlies the genetic code whose order is presented attending to the number of hydrogen bonds and to the chemical type. Algebraic operations are then introduced to reflect the quantitative relationships between codons. The principal aim of this work is to show that a simple algebraic structure of the genetic code can be defined on the sequence space allowing us to describe the mutations pathways in the molecular evolution process through the use of the endomorphisms $F: (Z_{64})^N \rightarrow (Z_{64})^N$, defined on the $Z_{64}$-modulo.

## 2. Theoretical Model

This paper rests on some well known algebraic structures: groups, rings, modules and algebras that we first built from the four DNA bases and from which we later developed the Genetic Code algebraic structures. A reminder of these algebraic structures appears in Appendix A. The theoretical abstract algebra used can be found in any textbook on this subject. We have specifically used textbooks by [7, 28, 21]. Thus, the genetic code algebraic structures have been defined according to the algebraic "classical" theory and extended to build the gene algebras. Actually, the gene algebras described here are DNA sequence algebras even though not all DNA sequences are genes. The term "gene" appears where we could generally use DNA sequence. Our first application, however, will be to study the molecular evolution process for which we would rather suggest a more compact denotation: "gene algebras".

*2.1. Four-base groups.*

Let the DNA bases be arranged according to their physico-chemical properties. The chemical types (purine and pyrimidine) and the number of hydrogen bonds are elements of the codon-anticodon interactions to obtain two orders in the base set. These elements are applied according to these criteria:

1) Chemical types cause the main difference between bases.
2) The greatest difference between one element and the next serve as criterion to select arrangements
3) The starting base needs a minimum hydrogen bond number.

As a result, the two orders {A, C, G, U} and {U, G, C, A} in these base set arrays made it possible to define a sum operation in several ways. From these, we obtained two cyclic groups isomorphic to $Z_4$ (group $Z_4$ of integers module 4). In Tables 1 A and B we show the sum tables of bases obtained from the two possible orders. Therefore, there are two cyclic groups: the primal and the dual group, corresponding to the ordered sets {A, C, G, U} and {U, G, C, A}.

Next, the two arrays {A, C, G, U} and {U, C, G, A} of the four-base set as well as the internal codon order reflecting the biological relevance of its base codons are considered to build two orders in the codon set. Our criterion of the maximum difference between bases must be initially applied to the third codon position. Next, it is applied to the first one and finally to the second base, i.e. from the less biologically relevant base to the most relevant base in the codon. We are able to establish the genetic code tables starting with the codon having the fewest hydrogen bonds. So, two possible codon set arrangements are induced. In the tree-entry tables, the order (0 to 63) is read in the following direction: the second base corresponding to 4 essential columns, the first base to 4 essential rows and the third base to sub-rows (Tables 2 and 3).

Table 1. Sum Operation tables defined in the set of four bases of the DNA. A: Primal algebra. B: Dual algebra.

| A | + | A | C | G | U | | B | + | U | G | C | A |
|---|---|---|---|---|---|---|---|---|---|---|---|---|
|   | A | A | C | G | U | |   | U | U | G | C | A |
|   | C | C | G | U | A | |   | G | G | C | A | U |
|   | G | G | U | A | C | |   | C | C | A | U | G |
|   | U | U | A | C | G | |   | A | A | U | G | C |

**Table 2.** The primal genetic code table induced by the primal order {*A, C, G, T*}. The bijection between the primal genetic code Abelian group with $Z_{64}$ is also shown in the table.

|   | A | | | C | | | G | | | U | | | |
|---|---|---|---|---|---|---|---|---|---|---|---|---|---|
|   | No | (1) | (2) | No | (1) | (2) | No | (1) | (2) | No | (1) | (2) | |
| A | 0 | AAA | K | 16 | ACA | T | 32 | AGA | R | 48 | AUA | I | A |
|   | 1 | AAC | N | 17 | ACC | T | 33 | AGC | S | 49 | AUC | I | C |
|   | 2 | AAG | K | 18 | ACG | T | 34 | AGG | R | 50 | AUG | M | G |
|   | 3 | AAU | N | 19 | ACU | T | 35 | AGU | S | 51 | AUU | I | U |
| C | 4 | CAA | Q | 20 | CCA | P | 36 | CGA | R | 52 | CUA | L | A |
|   | 5 | CAC | H | 21 | CCC | P | 37 | CGC | R | 53 | CUC | L | C |
|   | 6 | CAG | Q | 22 | CCG | P | 38 | CGG | R | 54 | CUG | L | G |
|   | 7 | CAU | H | 23 | CCU | P | 39 | CGU | R | 55 | CUU | L | U |
| G | 8 | GAA | E | 24 | GCA | A | 40 | GGA | G | 56 | GUA | V | A |
|   | 9 | GAC | D | 25 | GCC | A | 41 | GGC | G | 57 | GUC | V | C |
|   | 10 | GAG | E | 26 | GCG | A | 42 | GGG | G | 58 | GUG | V | G |
|   | 11 | GAU | D | 27 | GCU | A | 43 | GGU | G | 59 | GUU | V | U |
| U | 12 | UAA | - | 28 | UCA | S | 44 | UGA | - | 60 | UUA | L | A |
|   | 13 | UAC | Y | 29 | UCC | S | 45 | UGC | C | 61 | UUC | F | C |
|   | 14 | UAG | - | 30 | UCG | S | 46 | UGG | W | 62 | UUG | L | G |
|   | 15 | UAU | Y | 31 | UCU | S | 47 | UGU | C | 63 | UUU | F | U |

[1] The base triplets (codons).. [2] The one letter symbol of amino acids; "-" corresponds to stop codons.

**Table 3.** The dual genetic code table induced by the dual order {*U, G, C, A*}. The bijection between the dual genetic code Abelian group with $Z_{64}$ is also shown in the table.

|   | U | | | G | | | C | | | A | | | |
|---|---|---|---|---|---|---|---|---|---|---|---|---|---|
|   | No | (1) | (2) | No | (1) | (2) | No | (1) | (2) | No | (1) | (2) | |
| U | 0 | UUU | F | 16 | UGU | C | 32 | UCU | S | 48 | UAU | Y | U |
|   | 1 | UUG | L | 17 | UGG | W | 33 | UCG | S | 49 | UAG | - | G |
|   | 2 | UUC | F | 18 | UGC | C | 34 | UCC | S | 50 | UAC | Y | C |
|   | 3 | UUA | L | 19 | UGA | - | 35 | UCA | S | 51 | UAA | - | A |
| G | 4 | GUU | V | 20 | GGU | G | 36 | GCU | A | 52 | GAU | D | U |
|   | 5 | GUG | V | 21 | GGG | G | 37 | GCG | A | 53 | GAG | E | G |
|   | 6 | GUC | V | 22 | GGC | G | 38 | GCC | A | 54 | GAC | D | C |
|   | 7 | GUA | V | 23 | GGA | G | 39 | GCA | A | 55 | GAA | E | A |
| C | 8 | CUU | L | 24 | CGU | R | 40 | CCU | P | 56 | CAU | H | U |
|   | 9 | CUG | L | 25 | CGG | R | 41 | CCG | P | 57 | CAG | Q | G |
|   | 10 | CUC | L | 26 | CGC | R | 42 | CCC | P | 58 | CAC | H | C |
|   | 11 | CUA | L | 27 | CGA | R | 43 | CCA | P | 59 | CAA | Q | A |
| A | 12 | AUU | I | 28 | AGU | S | 44 | ACU | T | 60 | AAU | N | U |
|   | 13 | AUG | M | 29 | AGG | R | 45 | ACG | T | 61 | AAG | K | G |
|   | 14 | AUC | I | 30 | AGC | S | 46 | ACC | T | 62 | AAC | N | C |
|   | 15 | AUA | I | 31 | AGA | R | 47 | ACA | T | 63 | AAA | K | A |

[1] The base triplets (codons). [2] The one letter symbol of amino acids; "-" corresponds to stop codons.

*2.2. Codon groups and the $Z_{64}$-algebras of the genes*

Now, we can introduce the sum operation of codons *XYZ* and *X´Y´Z*. It is defined according to the classical group definition presented in Appendix A. Actually, if a cyclic group structure in the codon set is assumed, then its bijection -defined by the codon ordering- with $Z_{64}$ (the Abelian group $Z_{64}$ of integers module 64) causes such a unique definition of the codon sum that an isomorphism is evident. The cyclic character is desired because the genetic code should represent "an equilibrium status".

Alternatively, we give an adequate definition of "codon sum" regarding the distinction between the base positions in the codon, the order of the four bases set and the base sum operation. In this algorithm, the cyclic character of the sum of codons is hereditarily obtained from the base sum, while the order of importance of the bases is emphasized in the codons to establish the sum algorithm. The operation sum between two codons is obtained from the less biologically important base (third codon position: Z and Z´) to the most important base (second codon position: Y and Y´):

1) The corresponding bases in the third position are added according to the sum table.

2) If the resultant base of the sum operation is previous in order to the added bases (the orders in the set of bases), then the new value is written and the base C (or G if the dual group of bases is used) is added to the next position.

3) The other bases are added according to the sum table, step 2, going from the first base to the second base.

For instance, to perform the sum of codons AGC and UGU by using the primal group of bases (Table 1 A) we have:

C+U=A, the third bases are added and the base C is added to the next position because base A precedes bases C and U in the set of ordered bases {A, C, G, U}.

A+U+C=U+C=A, the first bases and the base C obtained in the first step are added. Again, base C is added to the next position.

G+G+C=A+C= C, the second bases are added to base C obtained in the second step. Finally, we have

AGC + UGU =ACA

The sum operation defined satisfies the sum group axioms (see Appendix A). As a result in the set of codons $C_g$ we will define two cyclic Abelian groups with operation sum "+" (denoted here as $(C_g, +)$): A primal codon group induced by the primal base group and a dual codon group induced by the dual base group. Since all finite cyclic groups with the same number of elements are isomorphic, then, groups $(C_g, +)$ are isomorphic to $(Z_{64}, +)$. So, for instance, in the primal algebra we can compute:

| AGC ↔ 33 | AGC ↔ 33 | CCC ↔ 21 |
|---|---|---|
| +UGU ↔ 47 | +AGA ↔ +32 | +AAU ↔ +3 |
| ACA ↔ 16 *mod* 64 | AAC ↔ 1 *mod* 64 | GCA ↔ 24 *mod* 64 |

Furthermore, this isomorphism allows us to refer to even or odd elements of $C_g$ (see Tables 2 and 3). That is to say, for instance, in Table 2 the codons with bases A and G in the thirst position are even codons and the codons with bases U and C are odds codons.

The product of element $k \in (Z_{64}, +)$ times element $XYZ \in (C_g, +)$ can be defined as an abbreviated sum, like in ordinary arithmetic, so that the symbol $k(XYZ)$ means the sum of $k$ times the term $XYZ$. Moreover, if for different $n$ natural numbers the product $nu$ exhausts the group, it is said that element $u \in C_g$ generates group $(C_g, +)$ and $u$ is called its generator. Generators of group $(C_g, +)$ are odd codons, also called units of $(C_g, +)$. On the whole, in one of the groups $(C_g, +)$, if codon $XYZ \in C_g$ is a generator of this group, then, for all codons $u \in C_g$ we can write $u = k(XYZ)$ where $k \in Z_{64}$. Specifically, for all $u \in (C_g, +)$ there exists $k \in (Z_{64}, +)$ such that $u = k(X_1Y_1Z_1)$, where $X_1Y_1Z_1 =$ AAC or $X_1Y_1Z_1 =$ UUG (see Table 2 and 3) Next, if we define the product operation (denoted by $\otimes$) in the set $C_g$ so that it is distributive with respect to the sum operation and the equality $(X_1Y_1Z_1 \otimes X_1Y_1Z_1) = X_1Y_1Z_1$, then, for all $u \in C_g$ and $v \in C_g$ we have:

$$u \otimes v = k(X_1Y_1Z_1) \otimes k'(X_1Y_1Z_1) = k \bullet k'(X_1Y_1Z_1 \otimes X_1Y_1Z_1) = k \bullet k'(X_1Y_1Z_1)$$

where "$\bullet$" denotes the product operation in the ring $(Z_{64}, +, \bullet)$, and ring $(C_g, +, \otimes)$ is isomorphic to ring $(Z_{64}, +, \bullet)$. Now, it evident that for all $x, y \in (Z_{64}, +, \bullet)$ and for all $u, v \in (C_g, +)$, the external law $f: Z_{64} \times (C_g, +) \to (C_g, +)$ given for $f(x,u) = xu = ux$ satisfy the R-Module and R-Algebra definitions, i.e. the group $(C_g, +)$ are cyclic $Z_{64}$-Modules and $Z_{64}$-Algebras defined over the ring $(Z_{64}, +, \bullet)$ [28]. In particular the external law $f$ can be considered as a module or an algebra endomorphism that include $f_k$ functions so that $f_k(u)=ku$ is an automorphism if, and only if, $k$ is an odd element of $Z_{64}$.

Next, these structures can be extended to the $N$-dimensional sequence space ($P$) consisting of the set of all $64^N$ DNA sequences with $N$ codons. Evidently, this set is isomorphic to the set of all $N$-tuples $(x_1,...,x_N)$ where $x_i \in C_g$. Then, set $P$ can be represented by all $N$-tuples $(x_1,...,x_N) \in (C_g)^N$ while algebraic group structures of $(C_g, +)$ can be extended to set $P$. As a result, groups $((C_g)^N, +)$ (primal and dual groups) will be the direct sum of $N$ groups $(C_g, +)$, that is:

$$(P, +) = ((C_g)^N, +) = (C_g, +) \oplus (C_g, +) \oplus ... \oplus (C_g, +) \quad (N \text{ times})$$

Similarly the $Z_{64}$-Modules and $Z_{64}$-Algebras of the groups $((C_g)^N, +)$ over the ring $(Z_{64}, +, \bullet)$ can be defined. At this point, for simplicity, we will denote these $Z_{64}$-algebras, $(C_g)^N$ and the ring $(Z_{64}, +, \bullet)$, $Z_{64}$. Next, in the algebras $(C_g)^N$, taken $X_1Y_1Z_1=AAC$ or $X_1Y_1Z_1=UUG$, the sets: $e_1=(X_1Y_1Z_1, 0,..,0)$, $e_2=(0, X_1Y_1Z_1,..,0)$,..., $e_n=(0, 0,.., X_1Y_1Z_1)$ are linearly independent, i.e. $\sum_{i=1}^{N} c_i e_i = 0$ implies $c_i=0$ for $i=1,...,N$, for any distinct $c_1, c_2, ...$, $c_N. \in Z_{64}$. Moreover, the representation of every DNA sequence $v \in (C_g)^N$ on the ring $Z_{64}$ as $v=x_1e_1+x_2e_2+...+x_ne_n$ ($x_i \in Z_{64}$) is unique. Then these generating sets are bases for the $(C_g)^N$ algebras [28]. It is said that elements $x_i \in Z_{64}$ are the coordinate of gene $v \in (C_g)^N$ in the canonical base ($e_1, e_2,..., e_n$) [21]. Next, for every endomorphism $f: (C_g)^N \to (C_g)^N$, the $N \times N$ matrix:

$$A = \begin{pmatrix} a_{11} & ... & a_{1n} \\ . & . & . \\ a_{n1} & ... & a_{nn} \end{pmatrix}$$

, whose rows are the image vectors $f(e_i)$ will be called the representing matrix of the endomorphism $f$, with respect to the base $e_i$. Since group $Z_{64}$ is an Abelian 2-group, according to the works of K.Shoda in 1928 [33], endomorphism $f: (C_g)^N \to (C_g)^N$ is an automorphism if, and only if, $\det(A)$ is not divisible by 2.

Likewise, algebraic structures for special genome sets are extended. Let $G$ be a set of genomes with $M$ DNA sequences (no coding regions included) then, it can be represented as the direct sum of $M$ algebras $P_i$, where $P_i$ could have different dimensions for $i=1,2,…M$, but for each $i$, the length of elements $P_i$ should be constant for all genomes in the set.

$$G = P_1 \oplus P_2 \oplus ... \oplus P_M$$

*2.3. Theoretical aspect of the gene algebras obtained.*

As groups $(C_g, +)$ are $Z_{64}$-algebras their elements are represented by elements of the ring $Z_{64}$. Then we will refer to even or odd elements of $(C_g, +)$. Moreover, because of the ring isomorphism: $(C_g, +, \otimes) \leftrightarrow (Z_{64}, +, \bullet)$, the algebraic properties of $(C_g, +, \otimes)$ are the same as for $(Z_{64}, +, \bullet)$.

The term "codon order" will be referred to as the "order of the codon" in a group structure. By definition, the order $n$ of an element $\alpha$ in the group $(C_g, +)$ is the least positive integer $n$ such that $n\alpha = 0$, where 0 is the neutral element for the sum. In particular, generators in a group have a maximal order. In the group $((C_g)^N, +)$ the order of element $\alpha=(\alpha_1, \alpha_2, ..., \alpha_N) \in (C_g)^N$ is the greatest order of their components $\alpha_i$ ($i=1,..N$).

Furthermore, elements of groups $(C_g, +)$ can be classified by their orders into seven classes. According to Lagrange theorem the orders of the elements in $Z_{64}$ are the divisors of 64, and these are the numbers with the form $2^m$ ($m = 0, 1, 2, 3, 4, 5, 6$). The elements with order $2^m$ have the form $2^{6-m} x$, where $x$ is an odd integer between 1 and $2^m-1$. We say that an element is $6-m$ high, of form $2^{6-m} x$ where $x$ is an odd integer. In Table 4 orders determine the partition of group $(Z_{64}, +)$ into seven classes. In $Z_{64}$ only, will elements with maximal order have their inverses for the operation "$\bullet$". In this table we also present odd elements and their inverses.

The set of mutant genes in respect to a wild type could be described through endomorphisms and automorphisms $f$ in $(C_g)^N$. Endomorphism $f: (C_g)^N \rightarrow (C_g)^N$ will be called local if there are $k \in \{1, 2,..., N\}$ and $a_{ik} \in Z_{64}$ ($i=1, 2,...,N$) such that $f(e_i) = (0,...,a_{ik},...,0) = a_{ik}e_i$.

**Table 4.** Partition of the group ($Z_{64}$, +) in seven classes. Elements with order 64 are presented in pairs in two columns, each one in the same line of its inverse for the operation "•" of the ring $Z_{64}$

| Order $2^m$ | 1 | 2 | 4 | 8 | 16 | 32 | 64[a] | |
|---|---|---|---|---|---|---|---|---|
| m | 0 | 1 | 2 | 3 | 4 | 5 | 6[a] | |
| Elements | 0 | 32 | 16 | 8 | 4 | 2 | 1 | 1 |
| | | | 48 | 24 | 12 | 6 | 3 | 43 |
| | | | | 40 | 20 | 10 | 5 | 13 |
| | | | | 56 | 28 | 14 | 7 | 55 |
| | | | | | 36 | 18 | 9 | 57 |
| | | | | | 44 | 22 | 11 | 35 |
| | | | | | 52 | 26 | 15 | 47 |
| | | | | | 60 | 30 | 17 | 49 |
| | | | | | | 34 | 19 | 27 |
| | | | | | | 38 | 21 | 61 |
| | | | | | | 42 | 23 | 39 |
| | | | | | | 46 | 25 | 41 |
| | | | | | | 50 | 29 | 53 |
| | | | | | | 54 | 31 | 31 |
| | | | | | | 58 | 33 | 33 |
| | | | | | | 62 | 37 | 45 |
| | | | | | | | 51 | 59 |
| | | | | | | | 63 | 63 |

.

This means that:

$$f(x_1, x_2, ... x_n) = (x_1, x_2, ... \sum_{i=1}^{n} x_i a_{ik}, ... x_n)$$

Therefore, any local endomorphism $F$ will be an automorphism if and only if element $a_{kk}$ in its representing matrix is an odd number. The determinant of this matrix is equal to $a_{kk}$.

The local endomorphism $f$ will be called diagonal if $f(e_k)=(0,...,a_{kk},...,0)=a_{kk}e_k$ and $f(e_i)=e_i$ for $i \neq k$. This means that:

$$f(x_1, x_2, ... x_n) = (x_1, x_2, ... a_{kk}x_k, ... x_n)$$

Likewise, a diagonal local endomorphism $f$ is an automorphism of $(C_g)^N$ if, and only if, $a_{kk}$ is an odd number. Finally, if the representing matrix with respect to the canonical base of endomorphism $f$ is a diagonal matrix, the endomorphism (automorphism) will be called diagonal. Thus it becomes clear that a diagonal endomorphism is a composition of $n$ diagonal local endomorphisms.

## 3. Results and Discussion

The nature of the codon-anticodon interactions allows us to explain the symmetry of the genetic code table [22]. In our description we obtained two codon arrays considering two important factors involved in these interactions: the hydrogen bond number and the chemical type. Thus, the algebraic symmetries in groups ($C_g$, +) are the result of the physico-chemical properties of four DNA bases included in the group definition. Consequently, such properties have to be closely connected to the physico-chemical properties of amino acids in groups ($C_g$, +)

*3.1. Preliminary Biological connections of the model.*

In the genetic code tables 2 and 3 there is an algebraic symmetry. For instance, in Table 2 the start codon AUG and the stop codon UAG are algebraically inverse, i.e. AUG + UAG = AAA. Similarly, in Table 3 we have: AUG + UAA = UUU. It is well known that codons AUG and UAG (UAA) are frequently used as the start codon and stop codon, respectively, in most of the genetic codes used by living organisms [26, 19] (http://www3.ncbi.nlm.nih.gov/Taxonomy/Utils/wprintgc.cgi?mode=t). Furthermore, in both tables, codons $X$U$Z$ coding to hydrophobic amino acids are algebraically inverse to codons $X'$A$Z'$ that codes to hydrophilic amino acids. The symmetry is also noted between the tables. It has been found that odd codons in Table 1 are even codons in Table 2 and that the unit element in Table 1 is number 63 in Table 2. Evidently the function φ: $X_1Y_1Z_1 \to X_2Y_2Z_2$ that turns codons $X_1Y_1Z_1$ from Table 1 into codons $X_2Y_2Z_2$ of Table 2 is the translation φ ($X_1Y_1Z_1$)=UUU - $X_1Y_1Z_1$. We can also see that, for instance, in both algebras when only two codons code to the same amino acid then, these codons have the same parity: they are either even or odd.

It is well known that single point mutations with the chemical type preserved (transitions) are frequently less dangerous than those that alter this property (transversions). Consequently, transitions are mostly single mutations found in nature [37]. In both genetic code tables (Table 1 and 2) it is found that transitions and transversions are connected with changes in the parity and the order of codons:

1. Without altering the codon parity, transitions in the second base can keep the codon order. For example, transitions: AAA<-->AGA (Lysine<-->Arginine); GAU<-->GGU (Aspartic<-->Serine). An extreme biological change is: UAG <--> UGG (Stop codon<-->Tryptophan).

2. Transition in either the first or the third bases keep the codon parity but the codon order may change. These mutations generally don't introduce extreme changes in the physico-chemical properties of amino acids.

3. Transversions in either the first or the second codon bases keep the codon parity but these transversions can vary the codon order. Transversions in the first base generally keep the hydrophobic properties, however, second codon transversions $X$UZ<--> $X$AZ are generally dangerous due to proteins.

4. Third base transversions change the codon parity and order; however, this kind of mutations generally does not produce extreme hydrophobic changes.

Evidently, from Tables 1 and 2, changes in the parity of codons (in a single point mutation) do not necessarily imply changes in the physico-chemical properties of amino acids. Additionally, few codon changes that preserve parity correspond to extreme changes in physico-chemical properties of amino acids. But, if most frequent mutations preserve the physico-chemical properties of the proteins, then, according to Table 1 and 2, most frequent mutations should keep parity.

The last observations are in harmony with the experimental data. The analysis of 749 aligned protease (pol) genes of HIV-1 has revealed that 90% of the 11172 mutations keep the parity of codons in respect to the wild type gene of the HXB2 clone. Moreover, mutations conferring resistance to drugs in the protease gene generally preserve the parity of codons (Table 5). Similarly, in mutations found in the human beta-globin gene variants, codon parities are preserved (Table 6).

*3.2. Endomorphisms and Automorphisms in the $Z_{64}$-Module $(C_g)^N$*

Very interesting connections between algebraic and biological properties in the $Z_{64}$-Module $(C_g)^N$ have been found considering the order of their elements (genes or DNA sequence). So, we would like to show that quantitative relationships between genes can be reflected in the use of the following theorem and its corollary.

**Table 5.** The mutations confering resistance to drugs in the protease gene. The wild type codons are from the sequence of the protease gene of the strain HXB2. The $Z_{64}$ value of the codons corresponds to the values of primal algebra (see Table 2). The mutations that increase the order are in bold type and those which alter the parity are in italic. The mutations in different combinations can increase the drugs resistance and many of them have cross-resistance.

| [a]Amino acid Mutations | Codon mutation | Value in $Z_{64}$ | Order $2^{m_i}$ change | $m_i$ change | Antiviral drug |
|---|---|---|---|---|---|
| A71I | GCU-->AUU | 27--->49 | 64--->64 | 6-->6 | ABT-378 |
| A71L | GCU-->CUC | 27--->53 | 64--->64 | 6-->6 | ABT-378 |
| A71T | GCU-->ACU | 27--->19 | 64--->64 | 6-->6 | Indinavir, Crixivan |
| A71V | GCU -->GUU | 27--->59 | 64--->64 | 6-->6 | Nelfinavir, Viracept |
| D30N | GAU-->AAU | 11--->3 | 64--->64 | 6-->6 | Nelfinavir, Viracept |
| *D60E* | *GAU-->GAA* | *11--->8* | *64-->32* | *5-->5* | DMP 450 PNU-140690 |
| G16E | GGG-->GAG | 42--->10 | 32--->32 | 5-->5 | ABT-378 |
| G48V | GGG -->GUG | 42--->58 | 32--->32 | 5-->5 | Telinavir MK-639 |
| G52S | GGU-->AGU | 43--->35 | 64--->64 | 6-->6 | AG1343 |
| G73S | GGU-->AGU | 43--->35 | 64--->64 | 6-->6 | AG1343 MK-639 |
| H69Y | CAU-->UAU | 7--->15 | 64--->64 | 6-->6 | Aluviran, Lopinavir |
| **I47V** | **AUA-->GUA** | **48--->56** | **4-->8** | **2-->3** | ABT-378, BILA 2185 BS |
| I50L | AUU-->CUU | 51--->55 | 64--->64 | 6-->6 | BMS 232632, VX-478 |
| I54L | AUC-->CUC | 49--->53 | 64--->64 | 6-->6 | ABT-378 |
| *I54M* | *AUU-->AUG* | *51--->50* | *64-->32* | *5-->6* | BILA 2185 BS, VX-478 |
| I54T | AUC-->ACC | 49--->17 | 64--->64 | 6-->6 | ABT-378 |
| I54V | AUC-->GUC | 49--->57 | 64--->64 | 6-->6 | ABT-378, MK-639 |
| I82T | AUC-->ACC | 49--->17 | 64--->64 | 6-->6 | A-77003 |
| **I84A** | **AUA-->GCA** | **48-->24** | **4--->8** | **2-->3** | BILA 1906 BS |
| **I84V** | **AUA-->GUA** | **48-->56** | **4--->8** | **2-->3** | Nelfinavir, Viracept |
| K20M | AAG -->AUG | 2-->50 | 32--->32 | 5-->5 | Indinavir, Crixivan |
| K20R | AAG-->AGG | 2-->34 | 32--->32 | 5-->5 | Indinavir, Crixivan |
| **K45I** | **AAA-->AUA** | **0--->48** | **1-->4** | **0-->2** | DMP-323 |
| **K55R** | **AAA-->AGA** | **0--->32** | **1-->2** | **0-->1** | AG1343 |
| L10F | CUC-->UUC | 53--->61 | 64--->64 | 6-->6 | Lopinavir, BILA 2185 BS |
| L10I | CUC-->AUC | 53--->49 | 64--->64 | 6-->6 | Indinavir, Crixivan |
| L10R | CUC-->CGC | 53-->37 | 64--->64 | 6-->6 | Indinavir, Crixivan |
| L10V | CUC-->GUC | 53-->57 | 64--->64 | 6-->6 | Indinavir, Crixivan |
| L10Y | CUC-->UAC | 53--->13 | 64--->64 | 6-->6 | BMS 232632 |
| L23I | CUA-->AUA | 52--->48 | 16-->4 | 4-->2 | BILA 2185 BS |
| L24I | UUA-->AUA | 60-->48 | 16--->4 | 4-->2 | Indinavir, Crixivan |
| L24V | UUA-->GUA | 60-->56 | 16--->8 | 4-->3 | Telinavir |
| *L33F* | *UUA-->UUC* | *60--->61* | *16-->64* | *4-->6* | ABT-538, BMS 232632 |
| L63P | CUC-->CCC | 53--->21 | 64--->64 | 6-->6 | ABT-378, AG1343 |
| L90M | UUG -->AUG | 62-->50 | 32--->32 | 5-->5 | Nelfinavir, Viracept |
| L97V | UUA-->GUA | 60--->56 | 16-->8 | 4-->3 | DMP-323 |
| M36I | AUG-->AUA | 50-->48 | 32--->4 | 5-->2 | Nelfinavir, Viracept |
| *M46F* | *AUG-->UUC* | *50-->61* | *32--->64* | *5-->6* | A-77009 |
| M46I | AUG-->AUA | 50-->48 | 32--->4 | 5-->2 | Indinavir, Nelfinavir |
| M46L | AUG-->UUG | 50-->62 | 32--->32 | 5-->5 | Indinavir, Crixivan |

Table 5. (continued)

| | | | | | |
|---|---|---|---|---|---|
| M46V | AUG-->GUG | 50-->58 | 32--->32 | 5-->5 | A-77006 |
| N88D | AAU-->GAU | 3--->11 | 64--->64 | 6-->6 | Nelfinavir, Viracept |
| N88S | AAU-->AGU | 3--->35 | 64--->64 | 6-->6 | BMS 232632, SC-55389A |
| P81T | CCU-->ACU | 23-->19 | 64--->64 | 6-->6 | Telinavir |
| R57K | AGA-->AAA | 32-->0 | 2-->1 | 1-->0 | AG1343 |
| R8K | CGA-->AAA | 36 -->0 | 16--->0 | 4-->0 | A-77003 |
| R8Q | CGA-->CAA | 36 -->4 | 16--->16 | 4-->4 | A-77004 |
| T91S | ACU-->UCU | 19-->31 | 64--->64 | 6-->6 | ABT-378 |
| V32I | GUA-->AUA | 56-->48 | 8--->4 | 3-->2 | A-77005,Telinavir |
| V75I | GUA-->AUA | 56-->48 | 8--->4 | 3-->2 | Telinavir |
| V77I | GUA-->AUA | 56-->48 | 8--->4 | 3-->2 | AG1343 |
| V82A | GUC-->GCC | 57-->25 | 64--->64 | 6-->6 | Ritonovir, Norvir |
| V82F | GUC-->UUC | 57-->61 | 64--->64 | 6-->6 | Ritonovir, Norvir, |
| V82I | GUC-->AUC | 57-->49 | 64--->64 | 6-->6 | A-77011 |
| V82S | GUC-->UCC | 57-->29 | 64--->64 | 6-->6 | Ritonovir, Norvir |
| V82T | GUC-->ACC | 57-->17 | 64--->64 | 6-->6 | Ritonovir, Norvir |

[a] All mutation informations contained in this printed table were taken from the Los Alamos web site: http://resdb.lanl.gov/Resist_DB. More details could be found in the Protease Mutations-by-Drug Map of the mentioned web site.

**Theorem**: For every pair $(\alpha, \beta)$ of elements of $(C_g)^N$, there is an endomorphism $f: (C_g)^N \to (C_g)^N$, such that $f(\alpha)=\beta$ if, and only if, the order of $\beta$ is a divisor of the order of $\alpha$.

Since $(C_g)^N$ is an homocyclic Abelian 2-group isomorphic to the homocyclic Abelian 2-group $(Z_{64})^N$ we will prove the theorem for this these special case.

*Proof*: It is well known that for any group homomorphism, every element $\alpha$ is carried out into an element whose order divides the $\alpha$ order. Then, the necessity is proved.

Now, let $\alpha \in (Z_{64})^N$ and $\beta \in (Z_{64})^N$ be of orders $2^{m_\alpha}$ and $2^{m_\beta}$, i.e. $\alpha = 2^{6-m_\alpha} x_\alpha$ and $\beta = 2^{6-m_\beta} x_\beta$, where $x_\alpha$ and $x_\beta$ are elements of maximum order, that is, of order 64. Starting with the element $x_\alpha$ we can complete a basis $(x_1, x_2 \ldots x_n)$ of $(Z_{64})^N$, where $x_1 = x_\alpha$. Analogously, starting with element $x_\beta$ a basis $(y_1, y_2 \ldots y_n)$ of $(Z_{64})^N$ can be completed, where $y_1 = x_\beta$. Later, there is a unique automorphism $f: (C_g)^N \to (C_g)^N$ such that $f(x_i) = y_i$ for $i = 1, 2, 3, \ldots N$. Then, if $2^{m_\alpha} \geq 2^{m_\beta}$, that is, $m_\alpha \geq m_\beta$, the endomorphism $2^{m_\alpha - m_\beta} f$ carries out the element $\alpha$ into the element $\beta$, as we wanted to prove. □

**Table 6**. Human hemoglobin variants caused by mutational events in the beta-globin gene. The $Z_{64}$ value of the codons corresponds to the values of primal algebra (see Table 2). Most of the mutations keep or decrease the codon order, then correspond to local endomorphisms. The mutations that increase the order are in bold type and those, which alter the parity, are in italic.

| [a]Amino acid Mutations | Codon mutation | Value in $Z_{64}$ | Order $2^{m_i}$ change | $m_i$ change | Biological effect | [b]PubMed ID |
|---|---|---|---|---|---|---|
| P36H | CCT->CAT | 23->7 | 64->64 | 6->6 | High oxigen affinity | 11939509 |
| T123I | ACC->ATC | 17->49 | 64->64 | 6->6 | asymptomatic | 11300351 |
| V20E | GTG->GAG | 58->10 | 32->32 | 5->5 | High oxigen affinity | 7914875 |
| V20M | GTG->ATG | 58->50 | 32->32 | 5->5 | High oxigen affinity | 7914875 |
| V126L | GTG->CTG | 58->54 | 32->32 | 5->5 | Neutral | 11939515 |
| V111F | GTC->TTC | 57->61 | 64->64 | 6->6 | Low oxygen affinity | 10975442 |
| H97Q | CAC->CAA | 5->4 | 64->16 | 6->4 | High oxigen affinity | 8571935 |
| V34F | GTC->TTC | 57->61 | 64->64 | 6->6 | High oxigen affinity | 10846826 |
| **E121Q** | **GAA->CAA** | **8->4** | **8->16** | **3->4** | | 8095930 |
| L114P | CTG->CCG | 54->22 | 32->32 | 5->5 | Non-functional | 11300352 |
| A128V | GCT->GTT | 27->59 | 64->64 | 6->6 | Mild instability | 11300349 |
| *H97Q* | *CAC->CAG* | *5->6* | *64->32* | *6->5* | High oxigen affinity | 8890707 |
| *D99E* | *GAT->GAA* | *11->8* | *64->8* | *6->3* | High oxigen affinity | 1814856 |
| D21N | GAT->AAT | 11->3 | 64->64 | 6->6 | | 8507722 |
| N139Y | AAT->TAT | 3->15 | 64->64 | 6->6 | High oxigen affinity | 8718692 |
| V34D | GTC->GAC | 57--9 | 64->64 | 6->6 | Unstable | 1260309 |
| E121K | GAA->AAA | 8->0 | 8->1 | 3->0 | | 7908281 |
| A140V | GCC->GTC | 25->57 | 64->64 | 6->6 | Mild polycythemia | 9028820 |
| K82E | AAG->GAG | 2->10 | 32->32 | 5->5 | Altered oxygen affinity | 9255613 |
| G83D | GGC->GAC | 41->9 | 64->64 | 6->6 | Hb Pyrgos (Normal) | 11843288 |
| D99N | GAT->AAT | 11->3 | 64->64 | 6->6 | High oxygen affinity | 1427427 |
| G15R | GGT->CGT | 43->39 | 64->64 | 6->6 | Neutral | 11939517 |
| V111L | GTC->CTC | 57->53 | 64->64 | 6->6 | Fannin-Lubbock variant | 7852084 |
| G119D | GGC->GAC | 41->9 | 64->64 | 6->6 | Fannin-Lubbock variant | 7852084 |
| H117Y | CAC->TAC | 5->13 | 64->64 | 6->6 | | 10870882 |
| E26K | GAG->AAG | 10->2 | 32->32 | 5->5 | | 9140717 |
| N108I | AAC->ATC | 1->49 | 64->64 | 6->6 | Low oxygen affinity | 12010673 |
| H146P | CAC->CCC | 5->21 | 64->64 | 6->6 | High oxygen affinity | 11475152 |
| H92Y | CAC->TAC | 5->13 | 64->64 | 6->6 | Cyanosis | 9494043 |

Table 6 (continued)

| | | | | | | |
|---|---|---|---|---|---|---|
| *C112W* | *TGT->TGG* | *47->46* | *64->32* | *6->5* | Silent and unstable | 8936462 |
| A111V | GCC->GTC | 25->57 | 64->64 | 6->6 | Silent | 7615398 |
| A123S | GCC->TCC | 25->29 | 64->64 | 6->6 | Silent | 7615398 |
| D52G | GAT->GGT | 11->43 | 64->64 | 6->6 | Silent | 9730366 |
| V126G | GTG->GGG | 58->42 | 32->32 | 5->5 | Mild beta-thalassemia | 1954392 |
| V67M | GTG->ATG | 58->50 | 32->32 | 5->5 | Severe instability | 8330974 |
| W15Stop | TGG->TAG | 46->14 | 32->32 | 5->5 | Beta-thalassemia | 10722110 |
| *F42L* | *TTT->TTG* | *63->62* | *64->32* | *6->5* | Hemolytic anemia | 11920235 |
| D99G | GAT->GGT | 11->43 | 64->64 | 6->6 | High-oxygen affinity | 9787331 |

[a] The amino acids are represented using one letter symbols. [b] The numbers are the PubMed ID of the articles indexed by MEDLINE and in which the mutations were reported.

Note that if the vectors α and β have the maximal order then, there is an automorphism $f$ such that $f(α)=β$. Codon usage tables for different living organisms suggest that codon usage is a warranty to always find odd codons in any natural gene (see Codon Usage Database: http://www.kazusa.or.jp/codon). Consequently, natural genes always have a maximum order. As a result, this theorem implies that, for every pair (α, β) of genes in the sequence space of dimension $N$, there is at least an automorphism such that $f(α)=β$. Moreover, since automorphism are one-one transformations on the group $(C_g)^N$, such that:

$$f(a \cdot (α+β)) = a \cdot f(α) + a \cdot f(β) \text{ for all genes } α \text{ and } β \text{ in } (C_g)^N \text{ and } a \in Z_{64}$$

Then, automorphisms forecast mutation reversions and if the molecular evolution process went by through automorphisms then, the observed current genes do not depend of the mutational pathway followed by the ancestral genes. In addition, the set of all automorphisms is a group. Additionally, according to the group theory, the set of all automorphisms $G$ of the group $(C_g)^N$ is a group. So, it could be possible to describe the mutation pathways in the sequence space by means of the group of their automorphisms. This means that if two different ancestral genes -in the molecular evolutions process- follow the mutational pathways through the automorphisms then algebraic differences between the ancestral genes are kept in actual descendant genes.

Next, as a particular case of the above theorem, the following corollary is obtained:

**Corollary**: A diagonal endomorphism $f: (C_g)^N \to (C_g)^N$, which transforms vector $\alpha$ to $\beta$ ($\alpha$, $\beta \in (C_g)^N$) exists if, and only if, for all pairs of coordinates $\alpha_i$, $\beta_i \in Z_{64}$ ($i=1, 2,...N$) of the vectors $\alpha$ and $\beta$, the inequalities $m_{\beta_i} \leq m_{\alpha_i}$ hold.

*Proof*: As a direct consequence of the theorem, for every $i=1,2,...N$, there exists a diagonal local endomorphism $f_i$ such that

$$f_i(\alpha_1, \alpha_2,...\alpha_i...\alpha_n) = (\alpha_1, \alpha_2,...\beta_i...\alpha_n)$$

But, this implies that there exists the diagonal endomorphism $f$ which is the compositions of $N$ endomorphisms $f_i$, such that $\beta = f(\alpha)$. Explicitly, if we have a number of $k$ pairs of components with $m_{i\beta} < m_{i\alpha}$ and $N-k$ with $m_{i\beta} = m_{i\alpha}$ then, $f$ is a composition of $k$ diagonal local endomorphisms and $N-k$ diagonal local automorphisms. □

According to this corollary, if $\alpha_i = 2^{6-m_{\alpha_i}} x_i$ and $\beta_i = 2^{6-m_{\beta_i}} y_i$ are the components of the vector $\alpha$ and $\beta$ respectively then, in the matrix representation $A$ of the endomorphism $f$ -on the basic generators of the module $(C_g)^N$- diagonal elements $a_{ii}$ are computed as.

$$a_{ii} = 2^{m_{\alpha_i} - m_{\beta_i}} y_i x_i^{-1}$$

In the experimental confrontation it was found that the 77,6% from the 749 mutants of HIV-1 protease genes have an order equal or smaller than the order of wild type HXB2. Hence, most of natural single mutations analyzed satisfy the corollary conditions, then to each one of them there exist a diagonal endomorphism that transform the clone HXB2 in that mutant. This percent increased when mutations that confer resistance to drugs were analyzed. The mutations were taken from Los Alamos HIV resistant database (http://resdb.lanl.gov/Resist_DB/default.htm). In Table 5 a set of mutations that confer resistance to drugs is shown. The values of codons correspond to values shown in the primal algebra. It can be seen that most of mutations respect to the wild type HXB2 protease gene satisfies the corollary. Similar situation is found in human hemoglobin variants caused by mutational events in the beta-globin gene (Table 6).

It can be observed that if for all pair of components $\alpha_i$ and $\beta_i$ the order is kept then, in the diagonal matrix will find only odd numbers and the endomorphism is an automorphism. Thus, if in one point mutation the codon order is kept then, there exist a local diagonal automorphism that transform the wild type into mutant. Moreover, in natural gene we can found manifold base substitutions in a gene keeping the codon order. In this case there is a diagonal automorphism to come from the wild type to the mutant gene. (see, for instance, the Explanation of Protease Mutations-by-Drug Map: http://resdb.lanl.gov/Resist_DB). All these suggest that diagonal endomorphisms are very frequents and automorphisms that keep the codon parity are frequents too. Curiously, it can be seen that the theorems and its corollary have been satisfied, in general, by the order presented in primal algebra.

*3.3. Automorphisms Representing Matrices between two Genes*

Since automorphisms will allows us to study the mutational pathway in the $N$-dimensional space of genes it will useful determine the representing matrix of any automorphism between two genes. Next, let $A$ be the matrix representation of the local endomorphism $f$: $(C_g)^N \to (C_g)^N$, on the canonical bases of a module $(C_g)^N$ (primal or dual), which transform components $\alpha_k$ ($k=1,..,N$) of the vector $\alpha$ into the component $\beta_l$ of the vector $\beta$ ($\alpha$, $\beta \in (C_g)^N$). Next, the action of the local endomorphism $f$ over $\alpha$ leads us to the equality:

$$\sum_{k=1}^{N} \alpha_k a_{kl} = \sum_{\substack{k=1 \\ k \neq i}}^{N} \alpha_k a_{kl} + \alpha_i a_{il} = \delta_i + \alpha_i a_{il} = \beta_l \ mod\ 64 \ (1 \leq i \leq N)$$

Subsequently, let $2^{m_{\alpha_i}}$ be the order of the component $\alpha_i$ of the vector $\alpha \in (C_g)^N$ and let $2^{m_{(\beta_l - \delta_i)}}$ be the order of the difference: $\beta_l - \sum_{\substack{k=1 \\ k \neq i}}^{n} \alpha_k a_{kl} = \beta_l - \delta_i$.

Then, the following proposition is deduced:

**Proposition**: Components $a_{il}$ of the column $l$ in matrix $A$ are computed as:

$$a_{il} = 2^{m_{\alpha_i} - m_{(\beta_l - \delta_i)}} y_{il} x_i^{-1} \ (i=1,2,..N)$$

where $x_i$, $y_{il} \in Z_{64}$ are odd numbers. What is more, if a component $\alpha_i$ of the vector $\alpha$ is fixed and taken as pivot then, components $a_{kl}$ of the column $l$ in the matrix $A$ can be arbitrarily chosen, for $k \neq i$, whenever $m_{\alpha_i} \geq m_{(\beta_l - \delta_i)}$.

*Proof.* Let us suppose that the matrix $A$ is the matrix representation of the local endomorphism $f$ on the canonical bases of an algebra $(C_g)^N$, in such a way that:

$$f(\alpha_1, \alpha_2, \ldots \alpha_i \ldots \alpha_N) = (\alpha_1, \alpha_2, \ldots \beta_l \ldots \alpha_N)$$

Since $\alpha_i = 2^{6-m_{\alpha i}} x_i$ and $\beta_l - \delta_i = 2^{6-m_{(\beta_l - \delta_i)}} y_{il}$, where $x_i$ and $y_{il}$ are odd numbers, then, the equality $2^{6-m_{\alpha_l}} x_i a_{il} = 2^{6-m_{(\beta_l - \delta_i)}} y_{il}$ means that $m_{\alpha_i} \geq m_{(\beta_l - \delta_i)}$. Because, in the Abelian group $Z_{64}$, for all elements $\alpha_i, \beta_l, \delta_l \in Z_{64}$, the equation $\delta_i + \alpha_i a_{il} = \beta_l$ has solution whenever $m_{\alpha_i} \geq m_{(\beta_l - \delta_i)}$ then, components $a_{il}$ are computed as:

$$a_{il} = 2^{m_{\alpha i} - m_{(\beta_l - \delta_i)}} y_{il} x_i^{-1} \quad (i=1,2,\ldots n)$$

In particular, according to the equation $\delta_i + \alpha_i a_{il} = \beta_l$, a component $\alpha_i$ of the vector $\alpha$ can be taken as pivot and matrix components $a_{kl}$, for $k \neq i$, can be arbitrarily selected whenever $m_{\alpha_i} \geq m_{(\beta_l - \delta_i)}$. □

This proposition allows us obtain the most simplified automorphism from a diagonal endomorphism. It is possible build any column of a matrix $A'$ with just two elements from a diagonal endomorphism matrix $A$, such that all their diagonal components will be odd numbers. We can consider, for example, the sequence α=*UAUAUGAGUGAC*. Let us suppose that, with successive mutations of this, it is turned out into the sequence β=*UGUAUAAGUCAG*. If the codons values are taken from the primal algebra (Table 2) these sequences correspond in the $Z_{64}$-algebra of $(C_g)^4$ to the vectors α=(15, 50, 35, 9) and β=(47, 48, 35, 6). As can be seen in the Table 4 for all vector components the inequalities $2^{m_{i\alpha}} \geq 2^{m_{i\beta}}$ hold. Hence, according to the corollary, there exists a diagonal endomorphism $f$, so that β=$f$(α). That is:

$$(47, 48, 35, 6) = (15, 50, 35, 9) \begin{pmatrix} 33 & 0 & 0 & 0 \\ 0 & 24 & 0 & 0 \\ 0 & 0 & 1 & 0 \\ 0 & 0 & 0 & 22 \end{pmatrix} (mod\ 64)$$

Next, columns 2 and 4 can be modified to obtain an automorphism using the proposition. It is convenient that components $a_{22}$ and $a_{44}$ will be odd numbers. So, we used $\alpha_2$=50 and $\alpha_4$=9 as pivot in the second and the fourth column, respectively, in order to obtain odd numbers in the diagonal. Finally, we have one of the possible variants:

$$(47, 48, 35, 6) = (15, 50, 35, 9) \begin{pmatrix} 33 & 34 & 0 & 0 \\ 0 & 1 & 0 & 0 \\ 0 & 0 & 1 & 31 \\ 0 & 0 & 0 & 1 \end{pmatrix} (mod\ 64)$$

This kind of automorphism could be useful to study the molecular evolution process. If the automorphism matrix $A$ is the most simplified between the wild type and mutant gene then, the component $a_{kl} \neq 0$ in the matrix $A$ could express the biological relationship between codons in the mutant positions $l$ and $k$. That is, for instance, in the set of HIV-1 protease mutant genes it is found the mutant: M46L/A71V/I84A. These mutations are: 50→62/27→59/48→24 (see the single point mutations in Table 5). This simultaneous mutations increase the resistant to the protease inhibitor BILA 1906 BS (520-fold resistant, see: http://204.121.6.61/Resist_DB/). By mean of the corollary and the proposition we will find the automorphism matrix that transforms the wild type HXB2 protease gene into this mutant. The matrix components different of cero are: $a_{ll}$=1 for $l \neq$ 46, 71, 84 and $a_{46\ 46}$=23, $a_{71\ 71}$= 33, $a_{84\ 84}$= 3 and $a_{71\ 84}$ =24. It can be verified that mutations A71V/I84A are associated resistant mutations (http://resdb.lanl.gov/Resist_DB/) i.e. the component $\alpha_{71}$ and $\alpha_{84}$ are biological and algebraically connected.

Automorphism representing matrices give us additional information about the mutational pathway in the $N$-dimensional space. The genetic fingerprint in the molecular evolution process could be expressed in these matrices.

*3.4. Stabilizer subgroup of the wild type conserved regions*

In this section we want consider the subset of automorphisms $Pr$ that for all gene $\alpha \in (C_g)^N$ keeps the parity of its components. It is not difficult to see that this subset is a subgroup of the set of all automorphisms $G$ of a group $(C_g)^N$. But this is even a big subgroup for practical purpose. It should be taken into account that in a wild type gene, normally, not all codon sequence is susceptible to experimental mutations. Usually, conserved, variables and hypervariables regions are found in genes. Next, let $S$ be the subset of mutant genes conserving the same regions from a wild type codon sequence $\alpha_0 \in (C_g)^N$. Then, according to the group theory [21], the set $St(\alpha_0)$ of automorphisms $f \in G$ that preserves these regions is a subgroup of $G$, that is:

$$St(\alpha_0) = \{f \in G, \text{ such that: } f(\alpha_0) = \beta \in S\} \subset G$$

This subgroup could be called the stabilizer subgroup in $G$ of the wild type conserved regions. For instance, in both the HIV protease and the human beta-globin genes, the whole mutants and their combinations (presented in the Tables 5 and 6) are included in their respective stabilizer subgroup. In general, the set $Pr \cap St(\alpha_0)$ is a subgroup of $G$. This subgroup could be the key to allow us reach more comprehension about the gene transformation pathways in the molecular evolution process. In Tables 5 and 6 we can see that the most of showed mutations, even its combinations, can be obtained by means of the automorphisms included in the subgroup $Pr \cap St(\alpha_0)$.

Automorphisms group is a warranty to keep the biological function of the genes, algebraically expressed keeping the codon parity and keeping or decreasing the codon order. Even though in the mutations that could affect level of biological activity the function is kept. This last case is present in the hemoglobin gene. In Table 6 it is seen that even those codon changes, which keep the hydrophilic and hydrophobic properties, altered the oxygen affinity making a clinical disorder in the patients, for instance: V20M**,** H97Q**,** D99E, V111F. But the biological function of the hemoglobin is kept, carrying oxygen by means of its hemo-group,

## 4. Conclusions

The $Z_{64}$-Modules and $Z_{64}$-Algebras of the genetic code are deduced from most elementals properties of the amino acids: the chemical type (purine and pirimidine), the number of hydrogen bonds and the biochemical distinction between the base positions in the codon. These structures of the genetic code should be consequence of biological relationship between codons usage, protein secondary structure and genes functions, mentioned in the introduction. As a result the genetic code Abelian groups induce an algebraic symmetry in the genetic code table and some connections between hydrophobic properties of coded amino acids and algebraic properties of the respective codons are found. For instance codons *XUZ* that codes to hydrophobic amino acids are the algebraic inverse of codons *X'AZ'* that codes to hydrophilic amino acids.

The natural extensions of these algebraic structures to the *N*-dimensional sequence space (*P*) consisting of the set of all $64^N$ DNA sequences with *N* codons lead us to two *N*-dimensional $Z_{64}$-Algebras. Experimental evidence show that biological pressure to keep a small change in the physico-chemical properties of the amino acids in the mutations leads to the preserving of the algebraic properties of codons. Then, it is no a surprise that mutation pathways followed by genes in the sequence space can be described by means of the automorphisms in the $Z_{64}$-Algebras of the DNA sequences. In general, the subgroup $Pr \cap St(\alpha_0)$ of the automorphism that preserve the codon parity and the conserved regions of the wild type gene are frequently observed in genes. This fact is found, for instance, in the mutations conferring resistance to drugs in the HIV protease gene and in the human hemoglobin variants caused by mutational events in the beta-globin gene. Then, the *N*-dimensional $Z_{64}$-Algebras.can helps us to study the molecular evolution process.

## Appendix A

Because the mathematical framework of this paper will be just about the abstract algebras, for the usefulness of the reader, in this appendix we remind the definitions of group, ring, endomorphism and automorphism [7, 28, 21]. Now, let S be a set.

**Definition:** A binary operation on $S$ is a function from $S \times S$ to $S$.

In other words a binary operation on $S$ is given when to every pair $(x, y)$ of elements of $S$ another element $z \in S$ is associated. If "•" is the binary operation on $S$, then $\bullet(x, y)$ will be denoted by $x \bullet y$, that is the image element $z$ is denoted by $x \bullet y$.

**Definition:** A group is the pair $(G, \bullet)$ composed by a set of elements $G$ and the binary operation $\bullet$ on $G$, which for all $x, y, z \in G$ satisfies the following laws:

  i. *Associative law*: $(x \bullet y) \bullet z = x \bullet (y \bullet z)$
  ii. *Identity law*: There exist in G a neutral element $e$ such that: $x \bullet e = e \bullet x$
  iii. *Inverse law*: For all element $x$ there is the symmetric element $x^{-1}$ respect to $e$ such that:
  $$x \bullet x^{-1} = x^{-1} \bullet x = e$$

In addition the group $(G, \bullet)$ is called an Abelian group (an additive group) if for all $x, y \in G$ the binary operation satisfy the commutative law: $x \bullet y = y \bullet x$. For the Abelian group the binary operation is denoted by the symbol "+" and is called sum operation. Now, the symbol 0 denotes the neutral element.

A set $G$ together with a binary operation is called a semi-group if satisfies the condition (i) of the above definition.

**Definition.** Let $(G, \bullet)$ be a group and let $n$ be the least positive integer, if it exists, such that $a^n = e$ then $n$ is called the order of $a$.

Here $a^n$ denote $a \bullet \ldots \bullet a$ ($n$ times). In the case of an Abelian group $(G, +)$ one writes $na$ instead of $a^n$, $a^n = a + \ldots \dot{a}$ ($n$ times).

**Definition**. A ring is a set $R$ with two binary operations, denoted by "+" and "•", with the following properties:

i. $(R, +)$ is a commutative group
ii. $(R, •)$ is a semi-group
iii. The following hold:

$$(x + y) • z = x • z + y • z$$
$$z • (x + y) = z • x + z • y.$$

A ring on the set $R$ is usually denoted by $(R, +, •)$. $(R, +, •)$ has a multiplicative identity if there is an element $1 = 1_R \in R$ such that for all $x \in R$, $x • 1 = 1 • x = x$. It is also said that $R$ is a ring with identity.

**Definition**. Let $R$ be rings with identity and let $M$ be an Abelian group. $M$ is called a $R$-Module is there exist an external law $f: R \times C_g \to C_g$, given for $f(x,u) = x\,u = u\,x$ that has, for all $x, y \in R$ and for all $u, v \in C_g$ the following properties:

1. $x\,(u + v) = xu + xv$
2. $(x + y)\,v = xv + yv$
3. $(x • y)\,v = x\,(yv)$
4. $1\,v = v$

If only $f(x,u) = xu$ ($f(x,u) = ux$) then is called a left $R$-Module (a right $R$-Module). Additionally, if the $R$-Module is also a ring and satisfies the properties:

$$x\,(uv) = (xu)v = u(xv)$$

Then, the $R$-Module is an algebra over the ring $R$ or a $R$-Algebra.

**Definition**. Let $(R, +, •)$ and $(S, \oplus, \otimes)$ be rings. Then a function $\varphi: (R, +, •) \to (S, \oplus, \otimes)$ is said to be a ring homomorphism if for all $x, y \in R$ the following hold:

i. $(x + y) = \varphi(x) \oplus \varphi(y)$

ii. $\varphi(x \bullet y) = \varphi(x) \otimes \varphi(y)$

If the homomorphism $\varphi$ is one-one then $\varphi$ is said to be a ring isomorphism. If $R=S$ then $\varphi$ is said to be a ring endomorphism and finally if $\varphi$ is one-one and $R=S$ then $\varphi$ is said to be a ring automorphism. The property *i* means that the function $\varphi$ is also a group homomorphism.

**Definition.** Let $M$ and $N$ be $R$-modules. A function $\varphi: M \to N$ is a homomorphism (i.e., an $R$-module homomorphism) provided it is a group homomorphism and if $x \in M$ and $r \in R$:

$$\varphi(xr) = \varphi(x)r$$

Similarly to rings we can tell about $R$-module isomorphism, endomorphism and automorphism.

**References**

1. Alf-Steinberger, C.: The genetic code and error transmission. Proc. Natl. Acad. Sci. USA, **64**, 584-591 (1969)

2. Balakrishnan, J.: Symmetry scheme for amino acid codons. Phys. Rev. E, **65**, 021912-5 (2002)

3. Bashford, J.D., Tsohantjis, I., Jarvis, P.D.: A supersymmetric model for the evolution of the genetic code. Proc. Natl. Acad. Sci. USA **95**, 987–992 (1998)

4. Bashford, J.D., Jarvis P.D.: The genetic code as a periodic table. Biosystems **57**, 147-61 (2000)

5. Beland, P., Allen, T.F.: The origin and evolution of the genetic code. J Theor Biol. **170**, 359-365 (1994)

6. Bertman, M.O., Jungck, J.R.: Group graph of the genetic code. J. Hered. **70**, 379-84 (1979)

7. Birkhoff, G., MacLane, S.: A survey of Modern Algebra. The Macmillan Company. New York (1941)

8. Crick, F.H.C.: The origin of the genetic code. J. Mol. Biol. **38**, 367-379 (1968)

9. Duret, L., Mouchiroud, D.: Expression pattern and, surprisingly, gene length, shape codon usage in *Caenorhabditis, Drosophila, and Arabidopsis.* Proc Natl Acad Sci **96**, 17–25 (1999)

10. Epstein, C. J.: Role of the amino-acid "code" and of selection for conformation in the evolution of proteins. Nature **210**, 25-28 (1966)

11. Friedman, S.M., Weinstein, I..B.: Lack of fidelity in the translation of ribopolynucleotides. Proc Natl Acad Sci USA **52**, 988-996 (1964)

12. Fuglsang, A.: Strong associations between gene function and codon usage. APMIS **111**, 843–7 (2003)

13. Gillis, D., Massar, S., Cerf, N.J., Rooman, M.: Optimality of the genetic code with respect to protein stability and amino acid frequencies. Genome Biology **2**, research0049.1–research0049.12 (2001)

14. Gu, W., Zhou, T., Ma, J., Sun, X., Lu, Z.: The relationship between synonymous codon usage and protein structure in Escherichia coli and Homo sapiens. Biosystems **73**, 89-97 (2004)

15. Gupta, S.K., Majumdar, S., Bhattacharya, K., Ghosh, T.C.: Studies on the relationships between synonymous codon usage and protein secondary structure. Biochem Biophys Res Comm **269**, 692-6 (2000)

16. He, M., Petoukhov, S.V., Ricci, P.E.: Genetic Code, Hamming Distance and Stochastic Matrices. Bull. Math. Biol. **66**, 1405-21 (2004)

17. Jiménez-Montaño, M.A., 1996. The hypercube structure of the genetic code explains conservative and non-conservative amino acid substitutions in vivo and in vitro. Biosystems **39**, 117-125.

18. Jiménez-Montaño, M.A., 1999. Protein Evolution Drives the Evolution of the Genetic Code and Vice Versa. BioSystems **54**: 47-64.

19. Jukes, T.H., Osawa, S.: Evolutionary changes in the genetic code. Comp Biochem. Physiol. B. **106**, 489-94 (1993)

20. Karasev, V.A., Stefanov, V.E.: Topological Nature of the Genetic Code. J. Theor. Biol. **209**, 303-317 (2001)

21. Kostrikin, A.I..: Introducción al algebra. Editorial MIR, Moscú (1980)

22. Lehmann, J.: Physico-chemical Constraints Connected with the Coding Properties of the Genetic System. J. theor. Biol. **202**, 129-144 (2000)

23. Makrides, S.C.: Strategies for achieving high-level expression of genes in Escherichia coli. Microbiol Rev **60**, 512–38 (1996)

24. Nakamura Y, Gojobori T, and Ikemura T. Codon usage tabulated from international DNA sequence database: status for the year 2000. Nucleic Acids Research **28**, pp 292 (2000)

25. Oresic. M., Shalloway, D.: Specific correlations between relative synonymous codon usage and protein secondary structure. J Mol. Biol. **281**, 31–48 (1998)

26. Osawa, S., Jukes, T.H., Watanabe, K., Muto, A.: Recent evidence for evolution of the genetic code. Microbiol Rev. **56**, 229-64 (1992)

27. Parker, J.: Errors and alternatives in reading the universal genetic code. Microbiol Rev. **53**, 273-298 (1989)

28. Redéi, L. Algebra, Vol.1. Akadémiai Kiadó, Budapest (1967)

29. Rose, G.D., Geselowitz A.R., Lesser G.J., Lee R.H., Zehfus M.H., Hydrophobicity of Amino Acids Residues in Globular Proteins. Science  229: 834-838. (1985)

30. Sánchez, R., Morgado, E., Grau, R.: The Genetic Code Boolean Lattice. MATCH Commun. Math. Comput. Chem **52**, 29-46 (2004)

31. Sánchez, R., Morgado, E., Grau, R.: A Genetic Code Boolean Structure. I. The Meaning of Boolean Deductions. Bull. Math. Biol. (article in press) doi:10.1016/j.bulm.2004.05.005 (2004)

32. Siemion, I.Z., Siemion, P.J., Krajewski, K.: Chou-Fasman conformational amino acid parameters and the genetic code. Biosystems 36, 231-238 (1995)

33. Shoda K.: Über die Automorphismen Einer Endlichen Abelichen Gruppe. Math. Ann. 100, 674-686 (1928)

34. Tao, X., Dafu, D.: The relationship between synonymous codon usage and protein structure. FEBS Lett **434**, 93–6 (1998)

35. Woese, C.R.: On the evolution of the genetic code. Proc Natl Acad Sci USA, **54**, 1546-1552 (1965)

36. Yang, C.M.: The naturally designed spherical symmetry in the genetic code. q-bio.BM/0309014 (http://arxiv.org/abs/q-bio.BM/0309014) (2003)

37. Yang, Z.: Adaptive molecular evolution.  In Handbook of statistical genetics, (Balding, M., Bishop, M. & Cannings, C., eds), Wiley:London, pp. 327-50 (2000)